\documentclass[11pt]{article}
\usepackage{multicol,citesort,epsfig}
\begin{document}

\title{Fluid-fluid transitions of hard spheres with a very short-range
attraction}

\author{{\bf Richard P. Sear}\\
~\\
Department of Physics, University of Surrey\\
Guildford, Surrey GU2 5XH, United Kingdom\\
email: r.sear@surrey.ac.uk}

\maketitle

\begin{abstract}
Hard spheres with an attraction of range a tenth to a hundredth
of the sphere diameter are constrained to remain fluid even
at densities when
monodisperse particles at equilibrium would have crystallised,
in order to compare with experimental systems which remain fluid.
They are found to have a fluid-fluid transition
at high density. As the range of the attraction tends to zero, the
density at the critical point tends towards the random-close-packing density
of hard spheres.
\end{abstract}


%\newpage
\begin{multicols}{2}

%intro

Argon forms a liquid because argon atoms attract
each other and these dispersion
attractions between the atoms are relatively long-ranged;
the volume over which one argon atom attracts
another is comparable to the volume one argon atom of the pair
excludes to another.
If we could reduce the range of the attraction between argon
atoms then the liquid phase would disappear from the equilibrium
phase diagram when the volume over which the atoms attract was
of order one tenth of the volume they exclude to each other.
Of course we cannot change the interaction between argon atoms
but there are well-established colloidal systems whose interactions
we can change. The liquid phase disappears from the equilibrium
phase diagram because the fluid-fluid transition is preempted
by the crystallisation of the fluid.
But although the fluid-fluid transition has disappeared
from the equilibrium
phase diagram of monodisperse particles, experiments
often do not observe crystallisation, presumably
due to a combination of a large
free energy barrier to crystallisation and the destabilising
effect of small amounts of polydispersity on the crystalline phase.
As crystallisation does not occur it does not preempt
the fluid-fluid transition, which is therefore observable.
With this in mind we study the behaviour of spherical
particles with a short-range attraction which are constrained to
remain fluid. We study attraction ranges
down to a hundredth of the diameter of the hard core
--- this is what we mean by very short-range attractions.
We find that as the range decreases, the density at the
critical point increases to very high values.
For a sufficiently short range the critical point lies
above the density of
the kinetic glass transition observed in experiments
on hard-sphere-like colloids.
Experiments on colloids with
very short-range attractions have found a threshold beyond which
diffusion ceases \cite{grant93,verduin95},
and Poon, Pirie and Pusey \cite{poon95} have previously suggested that
this is due to an arrested fluid-fluid phase separation.
If as we find, the dense fluid has a density above that of the
hard-sphere glass transition, then it is not surprising that
the dynamics of phase separation become arrested.

%why no crystal

Here we will not consider the crystalline phase at all. Our results
are for a system of particles which is constrained to remain fluid
at all temperatures and pressures; see refs.
\cite{debenedetti,penrose71,corti98}
for a discussion of the application of constraints to stabilise a
phase which would otherwise be metastable or unstable.
Although experiments on near-monodisperse colloidal spheres show that they
crystallise readily, at least
as long as the attraction is not too strong, polydisperse colloidal spheres
often never crystallise \cite{pusey91} and the presence of
a very short-range attraction makes the crystalline phase even
more sensitive to polydispersity \cite{sear99mp}.
By polydisperse spheres we mean that the
spherical particles do not all have the same diameter but have a range
of diameters. Our theory is a perturbation theory about a hard-sphere fluid
and so completely neglects the crystal. Thus we will not need
to explicitly apply a constraint within the theory.
We do, however, need to assume that it is possible to apply a constraint
to the system which has almost no effect on the fluid phase but completely
prevents crystallisation.

%potential

We chose a simple potential with a hard-sphere core and an attraction
in the form of a Yukawa function.
The hard-sphere+Yukawa potential is a spherically
symmetric pair potential so the interaction energy $v$ depends only on the
separation $r$ of the centres of the two particles,
\begin{equation}
v(r)=
\left\{
\begin{array}{ll}
\infty & ~~~~~~ r \le \sigma\\
-\epsilon\frac{\sigma}{r}\exp[\kappa(1-r/\sigma)] & ~~~~~~ \sigma<r\\
\end{array}\right. ,
\label{monoss}
\end{equation}
where $\sigma$ is the hard-sphere diameter and
$\epsilon$ is the energy of interaction for touching spheres.
With this potential the thermodynamic functions depend on the reduced
temperature $kT/\epsilon$, and the
reduced density $\eta=(N/V)(\pi/6)\sigma^3$ which is the
fraction of the volume occupied by the cores of the particles.
$k$, $T$, $N$ and $V$ are Boltzmann's constant, the temperature,
the number of particles and the volume, respectively.

%theory

We require a free energy for this potential which is accurate up
to very high densities, up to near random close packing which
is at a volume fraction $\eta\simeq0.64$-$0.65$
\cite{jodrey85,speedy98}. Speedy \cite{speedy98} has obtained,
from computer simulation data, an accurate equation of state of
hard spheres up to random close packing. This enables us to
use a perturbation theory, i.e., to start from the Helmholtz
free energy in the infinite
temperature limit of our model, which is hard spheres, and add
on the energy as a perturbation. Then our expression for the
Helmholtz free energy per particle $a$ at a temperature $T$ and a volume
fraction $\eta$ is
\begin{equation}
\beta a(\eta,T)=\beta a_{hs}(\eta)+\beta u(\eta,T),
\label{apert}
\end{equation}
where $a_{hs}$ is the Helmholtz free energy of hard spheres, $u$
is the energy per particle and $\beta=1/kT$. As the energy of a fluid
of hard spheres is zero, $\beta a_{hs} =-s_{hs}/k$, where $s_{hs}$ is the
entropy per particle
of hard spheres, which is, according to Speedy \cite{speedy98,note}
\begin{equation}
\frac{s_{hs}}{k}=1-\ln\rho+C\ln(\eta_0-\eta)+S_0+N^{-1}\ln N_g(\eta_0),
\label{speedy}
\end{equation}
where $C=2.8$, $S_0=-0.25$ and $N_g$ is
\begin{equation}
N_g(\eta_0)=\exp\left[N\left(\alpha-\gamma(\eta_0-\eta_m)^2\right)\right],
\label{ng}
\end{equation}
where $\alpha=2$, $\gamma=193$ and $\eta_m=0.555$. In these
equations the value of $\eta_0$ at any density
is determined by minimising the free energy at that density.
This form of the free energy is optimised for the dense fluid.
Essentially, if we start from any configuration of the dense
fluid and begin to expand all the spheres (so increasing the volume fraction)
then at some point the spheres will touch and then the spheres
cannot be expanded further. At this point the volume fraction is
$\eta_0$; this can be seen from the log term in eq. (\ref{speedy})
which diverges when $\eta=\eta_0$. If we start from different
configurations then after expansion of the spheres we may end
up with a different value of $\eta_0$.
The larger the difference
$\eta_0-\eta$ then the more room the spheres have which increases
the entropy. However, simulations show that there few
arrangements of the spheres
which have a large $\eta_0$, therefore there is an entropic cost
to being in an arrangement with a large $\eta_0$.
$N_g(\eta_0)$, eq. (\ref{ng}), is essentially
the number of ways of arranging spheres such that the maximum
possible volume fraction is $\eta_0$; it is maximal at $\eta_0=\eta_m$.
The competition between
the third and fifth terms in eq. (\ref{speedy}) then determines
the value of $\eta_0$. As the spheres touch when $\eta=\eta_0$ and
if we assume that the expansion is isotropic then the separation $b$
of spheres at a given $\eta$ and $\eta_0$ is
\begin{equation}
b/\sigma=(\eta_0/\eta)^{1/3},
\label{b}
\end{equation}
just as in a crystal.
The energy of attraction is approximated by the energy of interaction
of each sphere with its six neighbours \cite{jodrey85} at a separation $b$
\begin{equation}
u=-3v(b)=-3v((\eta_0/\eta)^{1/3}).
\label{u}
\end{equation}
As the energy depends on $\eta_0$, the total free energy, eq. (\ref{apert}),
is minimised to obtain $\eta_0$ at each density and temperature.

Guides to the accuracy of our free energy are obtained by comparison
with existing simulation data. For $\kappa=7$,
Hagen and Frenkel \cite{hagen94}
find a fluid-fluid critical point at $kT/\epsilon=0.41$, $\eta=0.26$,
whereas we predict $kT/\epsilon=0.54$, $\eta=0.30$. The agreement
is fair although not quantitative and we expect our theory to do
better at higher densities. Applying an approximation of the type,
eq. (\ref{u}), to a face-centred-cubic (fcc) crystal \cite{unpub} yields an
fcc-crystal--fcc-crystal critical point when $\kappa=100$ at
$kT/\epsilon=1.1$, $\eta=0.69$. Bolhuis, Hagen and Frenkel \cite{bolhuis94}
using computer simulation and perturbation theory predict
$kT/\epsilon=0.70$, $\eta=0.71$. Again there is fair
but not quantitative agreement

%results

\begin{figure}
\caption{
The fluid phase diagrams in the temperature-density plane, for
four different ranges.
The curves are, from left to right,
for inverse ranges $\kappa=7$, 20, 40 and 100.
In each case the curve encloses the fluid-fluid coexistence region
and the critical point is the highest point on the curve.
}
\label{figcoex}
\begin{center}
\epsfig{file=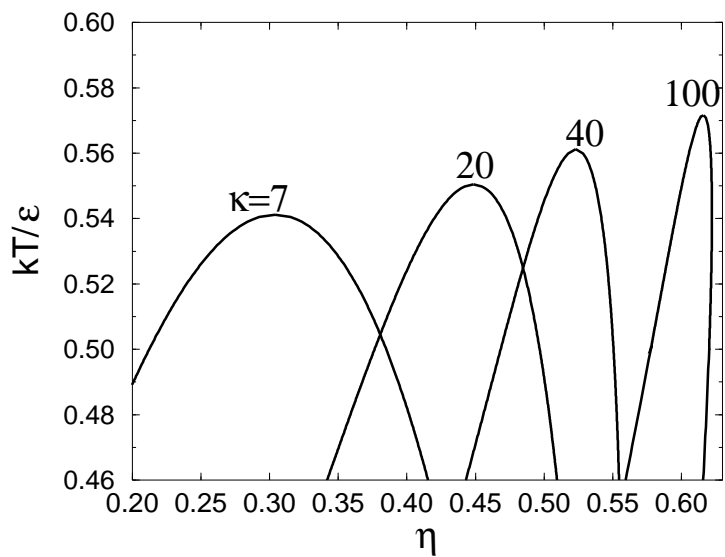,width=4.1in}
\end{center}
\end{figure}

Results for four, short, ranges are plotted in
fig. \ref{figcoex} \cite{note}. A simple liquid such as argon is reasonably
well modeled by an attraction of inverse range $\kappa=1.8$
\cite{henderson78}.
The results are for inverse ranges up to two orders of magnitude
greater. The notable feature is that the critical densities
and the densities of the liquid phase are high and move
to higher density as the range decreases.
At high density the particles are pushed together until they
are within range of the attraction. This occurs at separations
between the surfaces of the spheres
$b-\sigma={\cal O}(\sigma\kappa^{-1})$. With the particles just 
within range of the attraction there is a clear energetic
driving force towards phase separation: the fluid lowers its
energy at fixed overall density by some of the fluid condensing
into a dense fluid where all the spheres are well within the
range of the attraction of their nearest neighbours.
This is just what was observed by Bolhuis, Hagen and Frenkel
\cite{bolhuis94} in the fcc crystal. In the absence of the
crystalline phase, due to polydispersity perhaps, the transition
simply shifts over to a fluid-fluid transition and it occurs
at a lower density due to the fact that the random close-packing density
which is the maximum density of amorphous spheres is lower than the
maximum density of spheres achievable in an fcc crystal.
Because of the smaller number of neighbours in the dense fluid
as compared to the crystal the transition shifts
to a lower temperature but in both cases the critical temperature
varies little with changing range.

%comparison with expt

\begin{figure}
\caption{
The fluid phase diagrams in the $\tau$-density plane,
for two different ranges.
}
\label{figctau}
\begin{center}
\epsfig{file=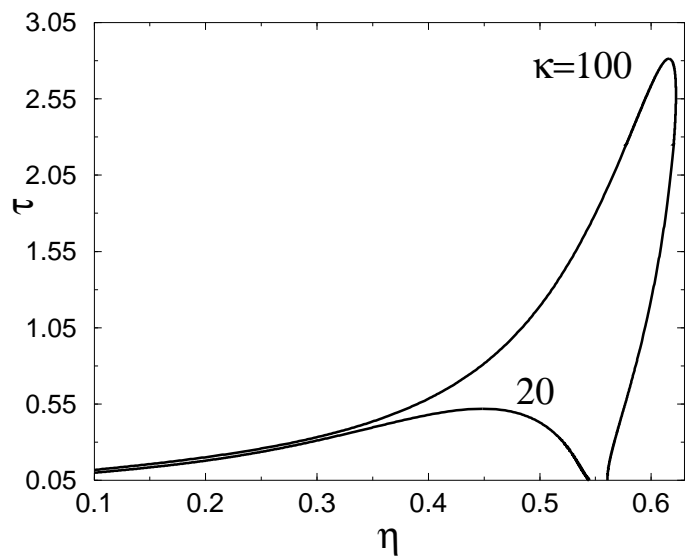,width=4.1in}
\end{center}
\end{figure}

Grant and Russel \cite{grant93}, and Verduin and Dhont \cite{verduin95}
have performed experiments on colloids which are hard-sphere-like
at high temperature but as the temperature is reduced the
solvent becomes a poor solvent for the alkane chains which are
grafted to the surface of the colloids. There is then a very short-range
attraction when the grafted layers of two colloids overlap.
Verduin and Dhont \cite{verduin95}
estimate the range of the attraction to be less than 1nm
for colloids with a diameter 80nm. They assess the strength of the
attraction via a parameter $\tau$ \cite{baxter68,stell91}
which is related to the second virial coefficient $B_2$ by
\begin{equation}
\tau=\frac{1}{4}\left(1-B_2/B_2^{hs}\right)^{-1},
\label{tau}
\end{equation}
where $B_2^{hs}$ is the second virial coefficient of hard spheres.
We have replotted the phase diagrams for $\kappa=20$ and 100
in the density-$\tau$ plane in
fig. \ref{figctau}. The experimental results are in the range
$\tau=0.1$ to 0.2, and $\eta=0.1$ to $\eta=0.4$. These densities and
temperatures lie within the coexistence region for $\kappa=100$.
They locate both a spinodal and a `static percolation' line
where diffusion of the particles stops.
At these densities and temperatures,
the fluid is unstable with respect to phase separation into a more
dilute phase and a very dense fluid phase --- this phase has a volume
fraction $\simeq0.56$.
Beyond a volume fraction of 0.56-0.58 \cite{pusey87}, the relaxation time
of a fluid of hard spheres exceeds typical experimental times which
are of the order of 1000s. This is referred to as a kinetic
glass transition, above it the samples are not equilibrated
on the experimental time scale and so are not an equilibrium fluid.
Thus as phase separation proceeds and
domains of this dense phase appear the phase separation dynamics
may become arrested due to the very slow relaxation within these
dense domains. This is the scenario suggested by Poon, Pirie and Pusey
on the basis of their experiments on colloids with a longer range
but still short-range attraction \cite{poon95}. Alternatively, as
Grant and Russel \cite{grant93} have suggested the phase separation may be
fluid-crystal phase separation.

%cf baxter

We predict a critical point at a density which increases as
the range decreases and at a value of $\tau$ which also increases
as the range decreases. Although we cannot perform calculations
at zero range, $\kappa=\infty$, extrapolation of our results
together with the results of Bolhuis, Hagen and Frenkel \cite{bolhuis94}
who were
able to study the zero range limit in the crystal, suggests that
in the zero-range limit there is a fluid-fluid critical point {\it at}
the random-close-packing density. The critical point would be
at roughly the same temperature as the critical points in
fig. \ref{figcoex}, which implies that as in the crystal
\cite{bolhuis94} it occurs at infinite $\tau$.
Baxter \cite{baxter68} solved the Percus-Yevick (PY) approximation
for hard-spheres with a
zero-range attraction. Within the PY approximation
there are 2 routes to the thermodynamic functions.
If the compressibility route is used the critical point is
at the low volume fraction
$\eta=0.12$ and at $\tau=0.098$ \cite{watts71}, whereas
via the energy route the prediction is $\eta=0.32$ and $\tau=0.12$
\cite{watts71}. Our results suggest that the critical point
predicted by the PY approximation may be an artifact of this
approximation.
Stell \cite{stell91} has shown that
the virial expansion is pathological for all finite $\tau$ in the limit
of the range of the attraction tending to zero. However, this
pathology originates in crystalline clusters which we have
eliminated with our constraint that crystalline configurations are not
allowed \cite{debenedetti,corti98}. So the pathological virial
expansion shows that in the zero-range limit,
the equilibrium, unconstrained, fluid is
completely unstable at finite $\tau$; see Refs.
\cite{bolhuis94,stell91,hemmer90,sear99} for the equlibrium phase
diagram in the zero-range limit.

%conclusion

In conclusion, we have determined the phase diagram of hard spheres
with an attraction with a range of order 0.1 or 0.01 of the hard-core
diameter, which are constrained not to crystallise. The fluid-fluid
transition persists, according to our approximate theory, for all
ranges of the attraction. As the range decreases
the density at the critical point increases and can become very high,
near the random-close-packing density of hard spheres. 
As the density is so high observing it will be difficult as the dynamics
are very slow at these densities; the densities can exceed that
of the glass transition of hard spheres. Due to these slow dynamics
a glass-glass transition may be observed instead of a fluid-fluid
transition but again it may be impossible to observe directly.
The difficulty in observing fully equilibrated coexistence does
not mean that the transition has no observable consequences.
Out of equilibrium systems tend to head toward equilibrium and
even if they do not reach equilibrium their final state may be,
roughly speaking, the point on the path to equilibrium where the
dynamics stop. This is our tentative interpretation of the
results on colloids with a very short-range
attraction \cite{grant93,verduin95}: the ceasing of diffusion
observed is arrested fluid-fluid phase separation.
One final point is that as the range decreases the critical point,
with its associated large fluctuations and critical slowing
down of the dynamics \cite{chaikin},
will approach the kinetic
glass transition. What effect this will have on the kinetic glass transition
is unknown.

%\newpage

\end{multicols}

\newpage

\begin{thebibliography}{99}
\bibitem{note} By construction, Speedy's entropy for hard spheres
\cite{speedy98} has a second-order phase transition when the value of $\eta_0$
obtained by maximising the entropy equals the
random-close-packing volume fraction. He assumes that beyond the
kinetic glass transition oberserved in experiment is an ideal
glass transition which is a true thermoynamic transition. It is
an open question whether this is correct or not. Our free energy
also has such a phase transition because it contains Speedy's
entropy. This transition occurs at densities above the fluid-fluid
transition and is not shown in our figures.
\end{thebibliography}
\end{document}